\documentclass[twocolumn,aps,prb,epsf,showpacs]{revtex4}
\usepackage{amsmath}
\usepackage{epsfig}
\usepackage{epsf}
\usepackage{array}

\begin{document}

\title{Nonequilibrium transport and optical properties of model metal--Mott-insulator--metal heterostructures}
\author{Satoshi Okamoto}
\altaffiliation[Electronic address: ]{okapon@ornl.gov}
\affiliation{Materials Science and Technology Division, Oak Ridge National Laboratory, Oak Ridge, Tennessee 37831, USA}
\date{\today }

\begin{abstract}
Electronic properties of heterostructures in which 
a finite number of Mott-insulator layers are sandwiched by semi-infinite metallic leads 
are investigated by using the dynamical-mean-field method combined with the Keldysh Green's function technique to account for 
the finite bias voltage between the leads. 
Current across the junction is computed as a function of bias voltage. 
Electron spectral functions in the interacting region are shown to evolve by an applied bias voltage. 
This effect is measurable by photoemission spectroscopy and scanning tunneling microscopy. 
Further predictions are made for the optical conductivity under a bias voltage as a possible tool to detect a deformed density of states. 
A general discussion of correlated-electron based heterostructures and future prospect is given. 
\end{abstract}

\maketitle


\section{Introduction}

Correlated-electron materials including transition-metal oxides have provided the basis for  
a variety of properties such as high-$T_c$ superconductivity in cuprates and colossal-magnetoresistance (CMR) in manganites.\cite{Imada98} 
Therefore it is anticipated that these materials will become the core of future electronic devices. 
For this purpose, the fabrication and characterization of interfaces including correlated-electron materials are of fundamental importance. 
In recent years, we have seen a tremendous amount of work on such heterostructures. 
Developments in thin film growth technique\cite{Kawasaki94} enable us 
to fabricate ``digital'' heterostructures out of correlated materials with atomic resolution,\cite{Ahn99,Izumi01,Gariglio02,Ohtomo02} 
and developments in scanning microscopy\cite{Batson93,Muller93,Browning93} enable us to characterize the physical properties 
including conduction band electron distribution of the heterostructures.\cite{Ohtomo02}
Furthermore, fabrication of practical devices has already started. 
This includes Josephson junction\cite{Bozovic03,Bozovic04} and tunneling magnetoresistance (TMR) junctions\cite{Sun96,Bowen03} 
with device characteristics that remain to be optimized. 

These experimental developments stimulated a great deal of theoretical research on correlated-electron heterostructures.%
\cite{Freericks04,Okamoto04a,Okamoto04b,Oka05,Lee06,Kancharla06} 
So far, most work has focused on the ground-state properties such as spectral function, charge density distribution, 
and linear conductance. 
As metal--semiconductor (or band insulator)--metal heterostructures provide one of the fundamental building brocks of 
current electronics,\cite{Esaki58,Josephson74,Giaever74} 
establishing the electric properties of correlated heterostructures is of great importance. 
Detailed analysis of the transport properties of correlated heterostructures including current-voltage characteristics 
is directly relevant to device applications.\cite{Bozovic03,Bozovic04,Sun96,Bowen03} 
However, this area remains largely unexplored. 

In this paper, we investigate the properties of heterojunctions consisting of a correlated insulator embedded in semi-infinite metallic leads. 
In particular, we focus on the steady state with an applied bias voltage between the leads. 
The effect of strong correlations is treated by using the dynamical-mean-field theory (DMFT)\cite{Georges96} 
combined with the Keldysh Green's function technique\cite{Keldysh65} to account for the finite bias voltage. 
We present current vs voltage characteristics of the junctions, position-dependent spectral functions, and 
optical conductivity spectra, and discuss how the transport properties of heterostructures are affected by correlations. 

The paper is organized as follows: 
In Sec.~\ref{sec:Model}, the theoretical model, formalism, and numerical techniques are outlined. 
In Sec.~\ref{sec:result}, we present numerical results, 
and finally in Sec.~\ref{sec:summary}, we discuss related work and future prospects. 

\section{Model and formalism}
\label{sec:Model}
\subsection{Model}
\begin{figure}[tbp]
\includegraphics[width=0.8\columnwidth,clip]{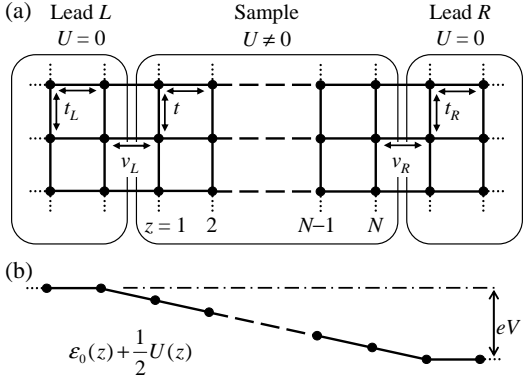}
\caption{Schematic view of the model heterostructure. 
(a) Projection on the [100] plane. System consists of a cubic lattice with discrete translational invariance in the $xy$ plane. 
Sample with a finite interaction $U$ ($1 \le z \le N$) couples to noninteracting leads $R$ ($z \ge N+1$) and $L$ ($z \le 0$). 
Transfer intensity in the sample and lead $R(L)$ are denoted by $t$ and $t_{R(L)}$, respectively. 
Hybridization strength between the sample and lead $R(L)$ is $v_{R(L)}$. 
(b) Potential profile across the junction. 
Note that potentials shown are $\varepsilon_0(z)$ shifted by $U(z)/2$.}
\label{fig:model}
\end{figure}

We consider electrons moving on a cubic lattice with lattice constant $a$ (=1) and discrete translational invariance in the $xy$ plane. 
Thus each site is labeled by $\vec r = (\vec r_\parallel,z)$ with $\vec r_\parallel = (x,y)$. 
A Hubbard-type interaction $U$ is introduced at a number $N$ of layers (sample $S$) located from $z=1$ to $N$, and 
noninteracting leads are located at $z \ge N+1$ (lead $R$) and $z \le 0$ (lead $L$). 
We consider the nearest-neighbor transfer $t$ ($t_\alpha$) of electrons in the sample (lead $\alpha$), 
the hybridization $v_\alpha$ between the sample and lead $\alpha$, 
and the layer-dependent potential $\varepsilon_0$ (see Fig.~\ref{fig:model}). 
Thus the Hamiltonian for this system is written as 
$H=H_S + \sum_{\alpha=R,L} H_\alpha + \sum_{\alpha=R,L} H_{S-\alpha}$ with  
\begin{equation}
H_S = - t \!\! \sum_{\langle \vec r, \vec r' \rangle, \sigma} \!\! \bigl( d_{\vec r \sigma}^\dag d_{\vec r' \sigma} + {\rm H.c.} \bigr)
+ U \sum_{\vec r} n_{\vec r \uparrow} n_{\vec r \downarrow} 
+ \sum_{\vec r, \sigma} \varepsilon_0 (\vec r) n_{\vec r \sigma} 
\label{eq:H_S}
\end{equation}
and
\begin{eqnarray}
H_\alpha \!\! &=& \!\! - t_\alpha \!\! \sum_{\langle \vec r, \vec r' \rangle, \sigma} \!\!
\bigl( d_{\vec r \sigma}^\dag d_{\vec r' \sigma} + {\rm H.c.} \bigr) 
+ \sum_{\vec r, \sigma} \varepsilon_0 (\vec r) n_{\vec r \sigma}, 
\label{eq:H_A} \\
H_{S-R} \!\! &=& \!\! - v_R \sum_{\vec r_\parallel, \sigma} 
\bigl( d_{\vec r_\parallel+N\hat z \, \sigma}^\dag d_{\vec r_\parallel+(N+1)\hat z \, \sigma} + {\rm H.c.} \bigr), 
\label{eq:H_S-R} \\
H_{S-L} \!\! &=& \!\! - v_L \sum_{\vec r_\parallel, \sigma} 
\bigl( d_{\vec r_\parallel \, \sigma}^\dag d_{\vec r_\parallel+ \hat z \, \sigma} + {\rm H.c.} \bigr), 
\label{eq:H_S-L} 
\end{eqnarray}
with
$H_S$ and $H_\alpha$ describing the interacting region and lead $\alpha$, respectively, and 
$H_{S-\alpha}$ the hybridization between the sample and the lead $\alpha$. 
$d_{\vec r \sigma}$ is an electron annihilation operator at position $\vec r$ with spin $\sigma$, and 
$n_{\vec r \sigma} = d_{\vec r \sigma}^\dag d_{\vec r \sigma}$. 
The position $\vec r$ in each term is constrained as explained above 
and $\hat z$ is the unit vector $\hat z = (0,0,1)$. 

\subsection{Layer dynamical-mean-field theory with finite bias voltage} 
To apply a finite bias voltage, 
we adiabatically turn on the hybridization and interaction between the leads and the sample.\cite{Caroli71,Datta} 
Before turning on the hybridization and interactions, 
the two leads have the chemical potentials $\mu_R$ and $\mu_L$ 
as well as the site potentials $\varepsilon_0(z \ge N+1) = \varepsilon_R$ and $\varepsilon_0(z \le 0) = \varepsilon_L$. 
The potential in the central region varies accordingly. 
We assume that the two leads are infinitely large and unaffected by the interactions, and thus remain in equilibrium. 
Under these conditions, 
the noninteracting leads are integrated out, and the electronic state of the interacting region may be described by the following 
Green's function matrices in 
the in-plane momentum $\vec k_\parallel$ and $z$-axis coordinate representation: 
\begin{eqnarray}
\hat G^{R} \bigl( \vec k_\parallel, \omega \bigr) \!\!&=&\!\! 
\Bigl[ \bigl\{ \hat g^{R} \bigl( \vec k_\parallel, \omega \bigr) \bigr\}^{-1} 
- \hat \Sigma^{R} \bigl( \vec k_\parallel, \omega \bigr) \Bigr]^{-1} , 
\label{eq:GR} \\
\hat G^K \bigl(\vec k_\parallel, \omega \bigr) \!\!&=&\!\! 
\hat G^R \bigl(\vec k_\parallel, \omega \bigr) \hat \Sigma^K \bigl(\vec k_\parallel, \omega \bigr) 
\hat G^A \bigl(\vec k_\parallel, \omega \bigr). 
\label{eq:GK}
\end{eqnarray}
Here, $R(A)$ and $K$ stand for retarded (advanced) and Keldysh components of the Green's function matrices, respectively, and 
$\hat G^A (\vec k_\parallel, \omega)= \bigl\{\hat G^R (\vec k_\parallel, \omega) \bigr\}^*$. 
$\hat g^{R}$ is the noninteracting retarded Green's function given by
$\bigl\{ \hat g^{R} \bigl( \vec k_\parallel, \omega \bigr) \bigr\}^{-1}
=
\omega+i0_+ - H_S \bigl( \vec k_\parallel, z,z'; U=0). 
$
$\hat \Sigma^R (\vec k_\parallel, \omega)$ and $\hat \Sigma^K (\vec k_\parallel, \omega)$ are the retarded and the Keldysh self-energies, respectively, 
representing
the effects of both electron correlation and the hybridization with the leads. 
The latter part of the self-energy can be obtained from the noninteracting Green's functions. 
Thus what must be determined is the self-energy due to correlations. 

In order to fix the correlation part of the self-energy, 
we generalize the layer DMFT.\cite{Potthoff99,Freericks01,Okamoto04b} 
Dynamical-mean-field theory using the Keldysh technique has been applied to solve the bulk correlated-electron model influenced by 
a time-dependent external field,\cite{Schmidt02,Freericks06} 
while our focus is the steady state. 
In the layer DMFT, 
the self-energy due to correlations is approximated to be diagonal in layer index and 
independent of in-plane momentum. 
Thus the lattice self-energy is written as 
\begin{eqnarray}
\Sigma_{z,z'}^\gamma \bigl(\vec k_\parallel, \omega \bigr) \!\!&\Rightarrow&\!\! \delta_{z,z'} \Bigl\{ \Sigma_z^\gamma (\omega) 
+ v_R^2 g_R^\gamma \bigl(\vec k_\parallel, \omega \bigr) \delta_{z,N} \nonumber \\
&& + v_L^2 g_L^\gamma \bigl(\vec k_\parallel, \omega \bigr) \delta_{z,1} \Bigr\}, 
\end{eqnarray}
where $\gamma =R,A$ and $K$. 
$g_\alpha^{R(A)} \bigl(\vec k_\parallel, \omega \bigr)$ is the retarded (advanced) Green's function of lead $\alpha=R,L$ 
projected on the layers adjacent to the interacting region, and 
\begin{equation}
g^K_\alpha \bigl(\vec k_\parallel, \omega \bigr) = \{1-2f_\alpha(\omega) \} 
\bigl\{ g_\alpha^R \bigl(\vec k_\parallel, \omega \bigr) - g_\alpha^A \bigl(\vec k_\parallel, \omega \bigr) \bigr\}
\end{equation}
is the Keldysh Green's function of lead $\alpha$. 
This function describes the distribution of electrons 
in terms of the Fermi distribution function $f_\alpha(\omega)= \{\exp \beta (\omega-\mu_\alpha)+1\}^{-1}$ 
with inverse temperature $\beta=1/T$ and chemical potential $\mu_\alpha$.

In DMFT, the quantum impurity model is introduced as a mathematical tool to compute the electron self-energy. 
The self-consistency condition of DMFT is closed 
by identifying the impurity Green's function $G_{imp,z}(\omega)$
with the local part of the lattice Green's function $G_{loc,z}(\omega)$ as 
\begin{eqnarray}
G_{imp,z}^\gamma (\omega) 
= G_{loc,z}^\gamma (\omega) \equiv \frac{1}{(2\pi)^2} \! \int \! (dk_\parallel)^2 G^\gamma_{zz} \bigl(\vec k_\parallel, \omega \bigr), 
\label{eq:Gloc}
\end{eqnarray}
where $\gamma =R(A)$ and $K$. 
The impurity model at each $z$ is characterized by the hybridization function $\Delta_z (\omega)$ and 
the nonequilibrium distribution function of electrons $f_{eff,z} (\omega)$ 
(in equilibrium, this is just the Fermi distribution function). 
The hybridization function represents the intersite virtual transfer of electrons in the form of the effective conduction band coupled to the impurity orbital. 
This includes intralayer and interlayer couplings and the hybridization between the sample and the leads. 
As in the equilibrium case,\cite{Georges96} 
the self-consistency condition determining this function is 
\begin{eqnarray}
\Delta_z^{R(A)} (\omega) = 
\omega - \varepsilon_0 (z) - \Sigma_z^{R(A)} (\omega)
- \Bigl\{ G_{loc,z}^{R(A)} (\omega) \Bigr\}^{-1} \!\!\!\!\! .
\label{eq:Delta_fix}
\end{eqnarray}
The nonequilibrium distribution function is fixed by the local Keldysh Green's function as 
\begin{equation}
G^K_{loc,z} (\omega) = \bigl\{ 1-2f_{eff,z} (\omega) \bigr\} 
\Bigl\{ G_{imp,z}^R (\omega) - G_{imp,z}^A (\omega) \Bigr\}. 
\label{eq:F_fix}
\end{equation}
Note that in the steady state we are focusing on, 
there is no net charge flow between the impurity orbital and the effective conduction band, i.e., 
the impurity model is in local ``equilibrium'' described by $f_{eff,z}$. 
Therefore, the Keldysh components of the self-energy and the hybridization function are related to the retarded ones by 
$\Sigma^K_z (\omega) = \bigl\{1-2f_{eff,z}(\omega) \bigr\} {\rm Im} \Sigma^R_z (\omega)$ and 
$\Delta^K_z (\omega) = \bigl\{1-2f_{eff,z}(\omega) \bigr\} {\rm Im} \Delta^R_z (\omega)$, respectively. 
Thus the self-consistency condition of DMFT is closed by 
Eqs.~(\ref{eq:Delta_fix}) and (\ref{eq:F_fix}) with Eq.~(\ref{eq:Gloc}).

The remaining task is to solve the quantum impurity model defined by the hybridization function $\Delta_z(\omega)$ and 
the distribution function $f_{eff,z}(\omega)$ with the local interaction. 
For this purpose, we employ the equation-of-motion decoupling (EOM) scheme,\cite{Appelbaum69,Lacroix81} 
which has been applied to study the nonequilibrium properties of quantum dots\cite{Meir91} 
and to solve the quantum impurity models of DMFT in equilibrium.\cite{Gros94,Jeschke05,Zhu04} 
In the EOM scheme, the retarded self-energy is given by 
(including spin dependence explicitly as $\sigma = \uparrow, \downarrow$ and $\bar \sigma = -\sigma$) 
\begin{equation}
\Sigma_{z \sigma}^R (\omega) = 
\frac{ U\{ \omega - \varepsilon_0(z) -  \Delta_{z \sigma}^R (\omega)\} \langle n_{z \bar \sigma} \rangle 
- U\Sigma_{1 z \sigma}^R (\omega)}
{\omega - \varepsilon_0(z) -  \Delta_{z \sigma}^R (\omega) - \Sigma_{2 z \sigma}^R (\omega) - 
U(1-\langle n_{z \bar \sigma} \rangle)},
\label{eq:Sigma}
\end{equation}
with
$\langle n_{z \sigma}\rangle 
= -\frac{1}{\pi} \!\int\! d \varepsilon \, f_{eff,z \sigma}(\varepsilon) \, {\rm Im}\,G_{imp,z \sigma}^R(\varepsilon)$, 
and 
\begin{eqnarray}
&& \hspace{-2em} 
\Sigma_{i z \sigma}^R(\omega) = - \frac{1}{\pi} \!\! \int \!\! d\varepsilon \, 
{\rm Im} \Delta_{z \bar \sigma}^R (\varepsilon)
A_{iz\sigma}(\varepsilon) \nonumber \\
&& \hspace{-2em} \times 
\biggl(
\frac{1}
{\omega +i0_+ + \varepsilon - 2 \varepsilon_0(z) - U} 
+ \frac{1}
{\omega +i0_+ - \varepsilon } \biggr), 
\label{eq:Sigma_Delta}
\end{eqnarray}
with
$A_{1z\sigma} (\varepsilon) = f_{eff,z\sigma} (\varepsilon)$ 
and $A_{2z\sigma} (\varepsilon) =1$.\cite{Zhu04} 

We pause to make one remark on solving the self-consistency equations. 
During the computation, we noticed that updating the distribution function directly from Eq.~(\ref{eq:F_fix}) 
by dividing both sides by $G_{imp,z}^R (\omega) - G_{imp,z}^A (\omega)$ 
does not provide smooth convergency. 
Instead, we found it more efficient to first rewrite the right hand side of Eq.~(\ref{eq:F_fix}) as 
$
2i \bigl\{1-2 f_{eff,z}(\omega)\bigr\} 
\bigl|G_{imp,z}^R (\omega) \bigr|^2
\bigl[ {\rm Im} \Sigma_z^R (\omega)+ {\rm Im} \Delta_z^R (\omega) \bigr] 
$
and then to update $f_{eff,z}(\omega)$ from  
$\bigl\{1-2 f_{eff,z}(\omega)\bigr\} {\rm Im} \Delta_z^R (\omega)$. 

It may be worth pointing out the limitation of the EOM scheme. 
As discussed in detail by Gros, the EOM scheme does not describe the correct metallic state of the Hubbard model at half filling. 
Thus, it fails to describe the bulk metal-insulator transition.\cite{Gros94} 
This is because the EOM scheme is essentially a strong coupling expansion, and therefore ``Kondo'' physics is not fully taken into account. 
Although some amount of Kondo physics is included by the exchange of electrons between the impurity orbital and the conduction band, 
it disappears in the particle-hole symmetric case. 
Therefore, it is favorable to take the on-site interaction $U$ larger than the critical value of the bulk Mott transition at half filling. 
When it is applied to the non-integer filling, the EOM scheme can reproduce the resonance-peak structure at the Fermi level at low temperature.\cite{Appelbaum69,Lacroix81} 
However, such Kondo physics is not fully taken into account. 
Therefore, results become less reliable at temperatures much lower than the characteristic Kondo temperature. 

Although it may not be easy, some improvement in the impurity solver of DMFT seems possible. 
For weak-to-intermediate interaction $U$, generalized iteration perturbation theory (IPT) \cite{Kajueter96,Schmidt02} seems realistic. 
But including higher-order perturbation processes in terms of $U$ may be necessary\cite{Fujii03} 
because the simple second-order perturbation with respect to $U$ does not capture the nonequilibrium Kondo feature when it is applied to 
the quantum dot problem.\cite{Hershfield91} 
For intermediate-to-strong coupling, the noncrossing approximation\cite{Pruschke93} may be applicable. 
However, this method is known to exhibit artificial structure in the spectral function at low temperature.\cite{Meir91} 
Another possibility is to use the recently developed continuous-time quantum Monte Carlo method,\cite{Rubtsov05,Werner06} 
which could be applied to a wider range of parameters. 
This method is formulated on the imaginary time axis assuming the system is in equilibrium. 
Therefore in order to apply it to the nonequilibrium situation, 
it is necessary to modify the method to deal with real time (or real frequency).

\subsection{Physical quantities}
Once self-consistency in the DMFT equations is achieved, one can proceed to compute physical quantities using the ``lattice'' Green's functions. 
Here, we first consider steady-state electric currents. 
In the steady state with a potential gradient along the $z$ direction, the current is uniform in space and time. 
Therefore one can measure the current through any bond along the $z$ direction (the steady current is conserved in the present formalism). 
Using the Keldysh Green's function matrix, 
the electric current $I$ per spin and per unit area between the $z$ and $z+1$ layers is computed as 
\begin{equation}
I = \frac{et}{2 \hbar} \! \int \! \frac{(dk_\parallel)^2 d \omega}{(2\pi)^3}
\Bigl\{ G_{z+1,z}^K \bigl(\vec k_\parallel,\omega \bigr) - G_{z,z+1}^K \bigl(\vec k_\parallel,\omega \bigr) \Bigr\}. 
\label{eq:Iz}
\end{equation} 

Another physical quantity of interest is the optical conductivity which can provide information about the dynamical properties of the system. 
In order to obtain the optical conductivity, a linear coupling between the current and the vector potential is first introduced 
according to the Pires-phase approximation as $\frac{1}{c} \vec I (t') \!\! \cdot \!\! \vec A (t')$. 
Then, the expectation value of the current $\langle I_l (t) \rangle$ to linear order in 
$A_{l'}(t')$ is computed using the nonequilibrium Green's functions. 
Here, $l,l'=x,y,z$, and $c$ is the velocity of light. 
The optical conductivity tensor $\sigma_{ll'} (\omega)$ is then obtained by $\sigma_{ll'} (\omega) = \langle I_l (\omega) \rangle/E_{l'}(\omega)$ 
where $\langle I_l (\omega) \rangle$ is a Fourier transform of $\langle I_l (t) \rangle$, and 
$E_{l'}(\omega)$ is an electric field given by $E_{l'}(\omega)=i \omega A_{l'}(\omega)/c$ with 
$A_{l'}(\omega)$ being a Fourier transform of  $A_{l'}(t)$. 
When the current and/or vector potential is along the $z$ direction, one has to subtract the contribution of the static current $I_z \ne 0$. 
When both the current and the vector potential are perpendicular to the $z$ direction, 
the optical conductivity is simply given by a Fourier transform of the current-current correlation function along the Keldysh contour. 
In this case, the conductivity $\sigma_{xx}$ is expressed using the interacting Green's functions as\cite{Lifshitz}
\begin{eqnarray}
\sigma_{xx} (\omega) 
\!&=&\! \frac{(et)^2}{\omega \hbar} 
\sum_{z, z'} \!\int\! \frac{(d k_\parallel)^2 d \omega'}{(2 \pi)^3} (2i \sin k_x)^2 \nonumber \\
&& \hspace{1em} \times \Bigl\{G^{--}_{z,z'} \bigl(\vec k_\parallel, \omega \bigr) G^{-+}_{z', z} \bigl(\vec k_\parallel, \omega+\omega' \bigr) 
\nonumber \\ 
&& \hspace{2em}
-G^{-+}_{z, z'} \bigl(\vec k_\parallel, \omega \bigr) G^{++}_{z', z} \bigl(\vec k_\parallel, \omega+\omega' \bigr) \Bigr\}, 
\label{eq:opcon}
\end{eqnarray}
where $\hat G^{--} = \{\hat G^K + \hat G^R + \hat G^A\}/2$, 
$\hat G^{-+} = \{\hat G^K - \hat G^R + \hat G^A\}/2$, and $\hat G^{++} = \{\hat G^K - \hat G^R - \hat G^A\}/2$. 
In Eq.~(\ref{eq:opcon}), we neglected the vertex correction which should become significant in finite dimension. 
Nevertheless, we believe that the essential features of the conductivity are captured by the interacting Green's functions. 
One can use a similar procedure to compute other dynamical quantities such as the dynamical spin susceptibility. 

\section{Results}
\label{sec:result}
In this section, we consider the particle-hole symmetric case when the bias voltage is absent 
to ensure the Mott insulating state in the interacting region. 
The chemical potentials and the site potentials of leads $R$ and $L$ are changed by a bias voltage $V$ as 
$\mu_L = \varepsilon_L = eV/2 = - \mu_R = -\varepsilon_R$. 
We assume that the potential in the central region varies linearly as 
$\varepsilon_0(z) = -U/2 + eV (N+1-2z)/(2N+2)$ interpolating $\varepsilon_L$ and $\varepsilon_R$. 
This assumption may be justified when the central region is highly insulating and the density profile is not changed by an applied bias voltage. 
A more realistic calculation requires the potential profile to be determined self-consistently by including long-ranged Coulomb repulsion. 
This is beyond the scope of the current study, but some discussion will be given later. 
We mainly use the parameters $U=15t$, $t_R=t_L=2.5t$, $v_R=v_L=t$, and $T=0.1t$ for the Hubbard heterostructures. 
Qualitative behavior does not depend on the choice of parameters 
unless the interaction parameter $U$ becomes as small as the bare bandwidth (and/or the interacting region becomes too thin)
and the interacting region becomes a ``metal'' even at half-filling. 

\begin{figure}[tbp]
\includegraphics[width=0.8\columnwidth,clip]{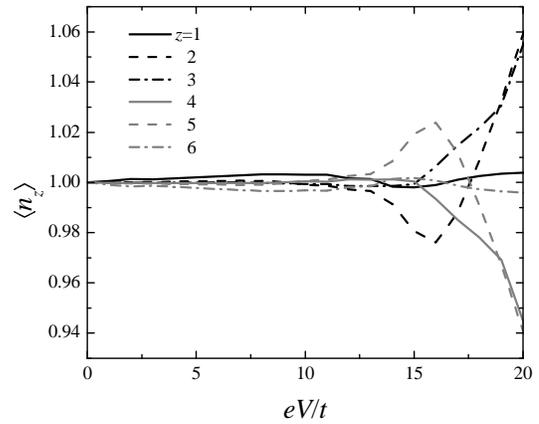}
\caption{Charge densities as functions of bias voltage for $N=6$ Hubbard heterostructure with $U=15t$.}
\label{fig:n_Hubb}
\end{figure}

Let us first check if the assumption of linear potential is reasonable or not. 
Figure \ref{fig:n_Hubb} plots the charge densities $\langle n_z \rangle = \sum_\sigma \langle n_{\sigma z} \rangle$ 
as a function of applied bias voltage for a $N=6$ heterostructure with $U=15t$. 
As can be seen, the charge densities are essentially unchanged up to $eV \sim 14t$ (deviation from $\langle n_z \rangle=1$ is smaller than 1\%; 
it is about 6\% even at high voltage $eV=20t$). 
Therefore, the assumption of a $z$-linear potential is rather realistic. 

\begin{figure}[tbp]
\includegraphics[width=0.8\columnwidth,clip]{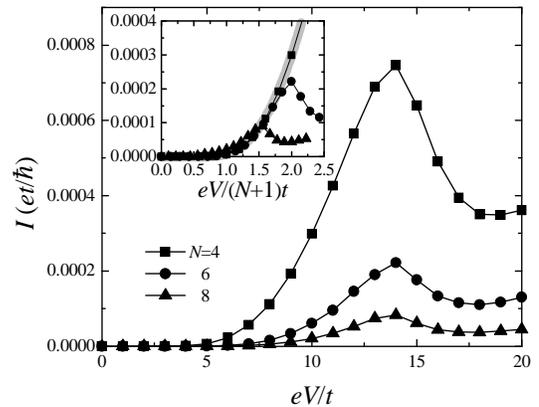}
\caption{Current-voltage characteristics of Hubbard heterostructures with $N=4,6$, and 8 with $U=15t$. 
Note that results shown are current per spin. 
Inset: current as a function of electric field defined by $E=eV/(N+1)$. 
The current at low voltage $eV/t<14$ falls onto the universal curve indicated by a light line $I=a_1 E \exp(-a_2/E)$ with $a_{1,2}$ fitting parameters. }
\label{fig:IV_Hubb}
\end{figure}

\begin{figure*}[tbp]
\includegraphics[width=1.4\columnwidth,clip]{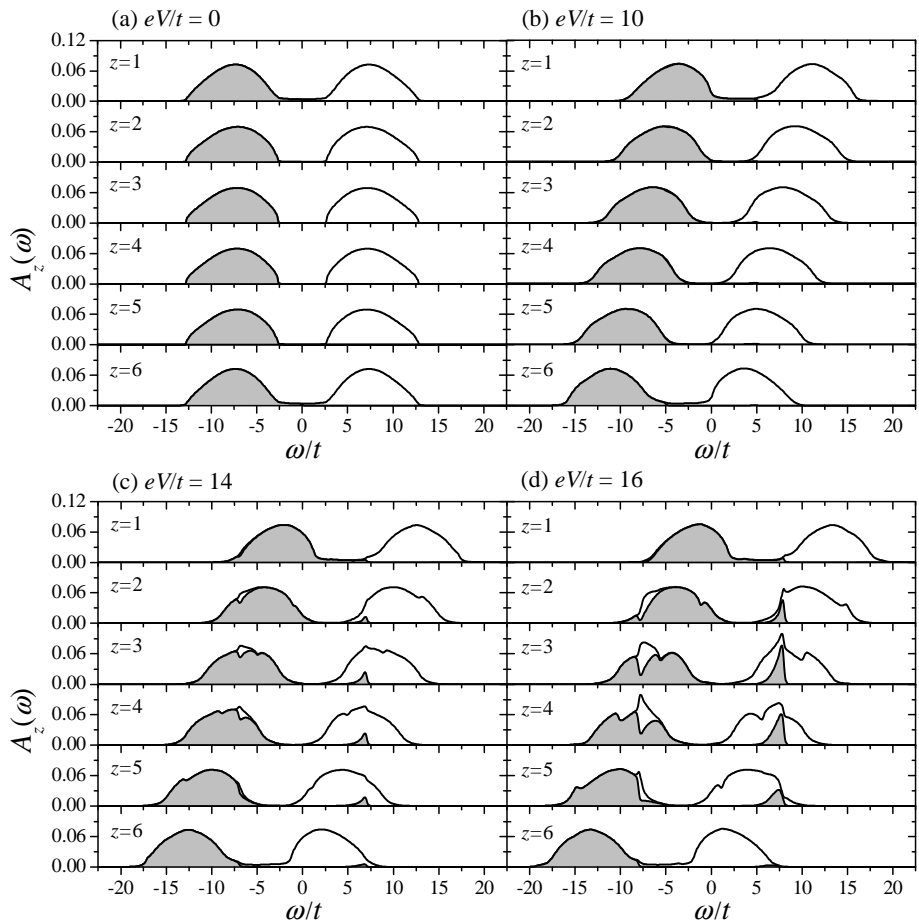}
\caption{Layer-resolved spectral function of electrons for $N=6$ Hubbard heterostructure with several choices of bias voltage indicated. 
On-site interaction is $U=15t$. 
Shaded area shows the region occupied by electrons, i.e., $f_{eff,z}(\omega) A_z(\omega)$. 
In (c) and (d), quasiparticlelike sharp structures can be seen in $A_z(\omega)$ at $\omega=\pm eV/2$ corresponding to the chemical potentials of leads $L$ and $R$. }
\label{fig:spectra_Hubb}
\end{figure*}

Next we discuss the current-voltage ($I$-$V$) characteristics of the Hubbard heterostructures. 
Numerical results are shown in Fig.~\ref{fig:IV_Hubb}. 
The thickness of the interaction region is changed as $N=4,6$, and 8. 
For small voltages $eV \alt 5t$, current is exponentially small. 
The current grows rapidly above $eV \sim 5t$, and it continues up to $eV \sim 14t$, above which it begins to decrease. 
Such a behavior is similar to that of a conventional metal--band-insulator--metal junction with the gap amplitude $\sim 5t$ 
and the distance from the middle of the gap to the top of the conduction band (or the bottom of the valence band) $14t$. 
Thus the $z$-axis transport is expected to be due to the interband Zener tunneling as discussed in Ref.~\onlinecite{Oka03}, in particular in the low voltage region. 
In Ref.~\onlinecite{Oka03}, breakdown of one-dimensional Mott insulators by an applied voltage is investigated using a finite-size Hubbard ring. 
The applied voltage is introduced by a time-dependent vector potential. 

This picture is further supported by replotting the current as a function of electric field $E=eV/(N+1)$. 
The result is shown in the inset of Fig.~\ref{fig:IV_Hubb}. 
As can be seen, the current at the low voltage region ($eV \alt 14t$) falls onto the universal curve given by $I=a_1 E \exp(-a_2/E)$ with $a_{1,2}$ fitting parameters 
as indicated by a light line. 
Similar results have been reported by Al-Hassanieh and co-workers who performed time-dependent density-matrix-renormalization group (DMRG) studies 
on one-dimensional heterostructures with a Mott-insulating region sandwiched by noninteracting leads.\cite{Al-Hassanieh07} 
This may indicate that the conventional band-insulator-like transport behavior is common for the Mott insulator in all dimensions. 
However, there has not been a direct observation confirming the Zener tunneling in correlated insulators. 
This is because the numerical methods used in the previous studies deal with the time evolution of the ground state under a time-dependent external field. 
So it is difficult to see the electron spectral function in the steady state. 
On the other hand, the present formalism directly deals with the steady state. 

\begin{figure}[tbp]
\includegraphics[width=0.8\columnwidth,clip]{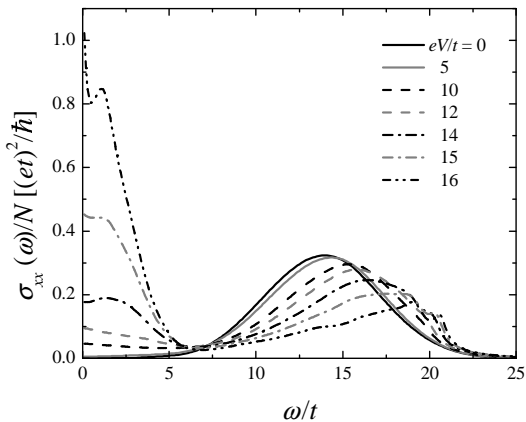}
\caption{Optical conductivity spectra of $N=6$ Hubbard heterostructure with several choices of bias voltage indicated. 
On-site interaction $U=15t$. 
}
\label{fig:opcon_Hubb}
\end{figure}

Figure \ref{fig:spectra_Hubb} shows the layer-resolved spectral functions $A_z(\omega) = -\frac{1}{\pi} {\rm Im} G_{loc,z}(\omega)$ 
for $N=6$ Hubbard heterostructure with several choices of bias voltage. 
Other parameters are the same as in Figs.~\ref{fig:n_Hubb} and \ref{fig:IV_Hubb}. 
Occupied regions, i.e., $f_{eff,z} A_z(\omega)$, are shaded. 
Figures~\ref{fig:spectra_Hubb} (a)-- \ref{fig:spectra_Hubb}(c) confirm the naive interpretation based on the conventional band insulator; 
about $5t$ of Mott gap is evident from (a); 
a chemical potential of lead $L$ ($eV=5t$) is located inside the upper Hubbard bands (b), 
and touches the top of the upper Hubbard band of the layer at $z=6$ at $eV \sim 14t$ (c). 
Above $eV \sim 14t$ (d), another tunneling process sets in from the top of the upper Hubbard band for layers $z<6$ to the unoccupied state of lead $R$, 
resulting in a negative differential conductance. 
At low bias voltage $eV \alt 14t$, the deformation of spectral functions is very weak 
because the amount of electrons injected in the upper Hubbard band 
(and holes injected in the lower Hubbard band) is too small (see also the voltage dependent charge densities in Fig.~\ref{fig:n_Hubb}). 
Therefore the magnitude of the gap controls the $z$-axis transport which is less sensitive to the detail of the model. 
This confirms the Zener tunneling mechanism. 

In addition to the ``rigid'' shift of the layer-dependent spectral functions, 
quasiparticle-like sharp features are visible in the spectral functions at $\omega \sim \pm eV/2$  
in the high bias voltage region $eV \agt 14t$ as seen in Figs.~\ref{fig:spectra_Hubb} (c) and \ref{fig:spectra_Hubb} (d). 
In the equilibrium case, such quasiparticle features are observed only at the Fermi level. 
In the current nonequilibrium case, it is possible to have two ``quasiparticle'' peaks at $\omega \sim \pm eV/2 = \mu_L,\mu_R$ 
as in the nonequilibrium quantum dot.\cite{Meir91} 
The existence of quasiparticles may indicate metallic transport properties, i.e., larger current densities. 
However, with the parameters used here, 
quasiparticle features appear at rather high voltage and the additional tunneling process sets in as discussed above. 
Therefore the current density is not enhanced. 
It would be very interesting to investigate the nonequilibrium behavior of 
correlated heterostructures with a smaller parameter $U$ (close to the critical value for the bulk Mott transition). 
If the ``quasiparticle'' structures appear before the chemical potential $\mu_L$ exceeds the upper edge of the spectral function at the rightmost layer, 
it would induce a larger current, resulting in highly nonlinear $I$-$V$ characteristics. 
Such a study requires impurity solvers suitable for weaker interactions. 

Layer-resolved spectral functions deformed by an applied bias voltage are essentially measurable by photoemission spectroscopy (PES) 
and scanning tunneling microscopy (STM). 
Yet, in light of the available spatial resolution, the latter seems plausible. 
Since, a current is already injected by an applied voltage, the insulating nature of the sample would not be a problem for STM. 
Although indirect, optical conductivity measurements might also be useful to investigate deformed spectral functions. 
Using Eq.~(\ref{eq:opcon}), in-plane optical conductivity spectra $\sigma_{xx}(\omega)$ are computed for $N=6$ Hubbard heterostructure 
with several choices of bias voltage as shown in Fig.~\ref{fig:opcon_Hubb}. 
Other parameters are the same as in Figs.~\ref{fig:n_Hubb}--\ref{fig:spectra_Hubb}. 
At low bias voltage, the spectrum is dominated by an inter-Hubbard-band transition appearing at $\omega \sim U=15t$. 
With increasing bias voltage, the peak position of the inter-Hubbard-band transition shifts upwards 
because of the change in the potential profile in the interacting region. 
The weight of the inter-Hubbard-band transition is reduced and transferred to the low frequency region. 
The Drude-like low frequency structure is rather broad until sharp resonancelike structures become clear in the spectral functions. 
In the actual optical conductivity measurement, one needs a rather thick interacting region compared with the radius of the incident light. 
Otherwise, the huge Drude response from the noninteracting leads would hide the low frequency features coming from the interacting region. 

\begin{figure*}[tbp]
\includegraphics[width=1.4\columnwidth,clip]{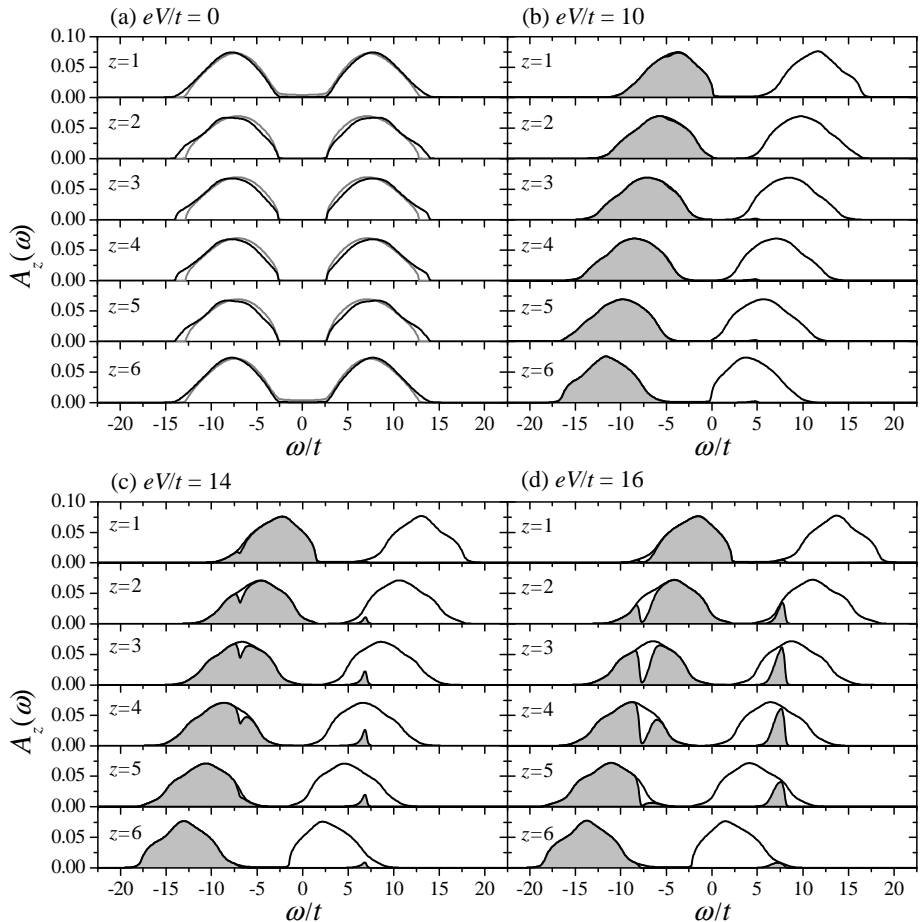}
\caption{Layer-resolved spectral functions of itinerant $c$ fermions for $N=6$ Falicov-Kimball heterostructure with several choices of bias voltage indicated. 
Parameters are the on-site interaction $U_{FK}=15 t$ and hopping amplitude $t_{FK}=1.5t$. 
Shaded area shows the region occupied by the $c$ fermions.
For comparison, spectral functions of $N=6$ Hubbard-model heterostructure with $U=15t$ and $eV=0$ are also shown as light lines in (a). 
Unlike Hubbard heterostructure, a quasiparticlelike sharp structure does not appear in $A_z(\omega)$ at any bias voltage.}
\label{fig:spectra_FK}
\end{figure*}

\subsection*{Comparison with Falicov-Kimball model}
In order to see the effect of the change in spectral functions, in particular the evolution of quasiparticle peaks, on the current-voltage characteristics, 
we have also applied the current DMFT method to the spinless Falicov-Kimball (FK) model. 
In this model, itinerant $c$ fermions described by  
$-t_{FK}\sum_{\langle \vec r, \vec r' \rangle} \bigl(c^\dag_{\vec r} c_{\vec r'}+ {\rm H.c.} \bigr)$
interact 
with localized $f$ fermions via the on-site Coulomb interaction $U_{FK} \sum_{\vec r} c_{\vec r}^\dag c_{\vec r} f_{\vec r}^\dag f_{\vec r}$.\cite{Falicov69} 
It is known that, when $\langle n^f_z \rangle = \langle f_z^\dag f_z \rangle \ne 0,1$, 
the FK model does not exhibit a quasiparticle feature unlike the Hubbard model.\cite{Si92} 
We take the local interaction as $U_{FK}=15t$ and the transfer intensity of $c$ fermions as $t_{FK}=1.5 t$ so that 
the positions of the upper- and lower-Hubbard bands and the magnitude of Mott gap become similar to those of the Hubbard model with $U=15t$ 
studied in the previous section. 

The $c$-fermion self-energy of the FK model has the same form as the electron self-energy of the Hubbard model $\Sigma_{z \sigma}^R (\omega)$ 
given in Eq.~(\ref{eq:Sigma_Delta}) with 
$\Sigma_{1(2) z \sigma}(\omega)=0$ and $\langle n_{z \bar \sigma} \rangle$ replaced with $\langle n^f_z \rangle$. 
For the computation, we fixed the mean density of $f$ fermions at each layer as $\langle n^f_z \rangle= 0.5$ for all applied bias voltage. 

\begin{figure}[tbp]
\includegraphics[width=0.8\columnwidth,clip]{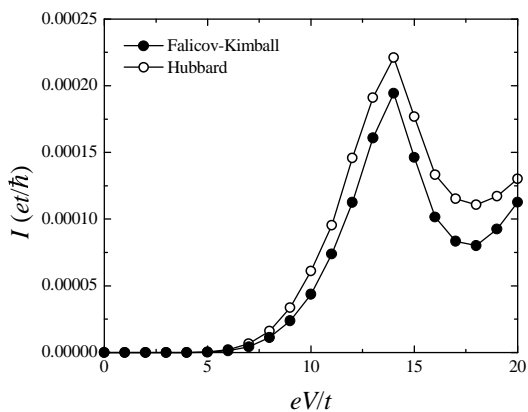}
\caption{Current-voltage characteristics of $N=6$ Falicov-Kimball heterostructure with $U_{FK}=15t$ and $t_{FK}=1.5t$. 
The electric current is carried by itinerant $c$ fermion. 
For comparison, the current per spin as a function of bias voltage of $N=6$ Hubbard heterostructure with $U=15t$ is also shown. }
\label{fig:IV_FK}
\end{figure}

Numerical results for the layer-resolved spectral functions of itinerant $c$ fermions for $N=6$ FK-model heterostructure are shown in Fig.~\ref{fig:spectra_FK} 
with several choices of bias voltage indicated. 
As seen in Fig.~\ref{fig:spectra_FK} (a), 
the inter-Hubbard-band peak-to-peak distance and the gap amplitude of the FK heterostructure are almost identical to those of the Hubbard heterostructure. 
It is also evident that the spectral functions of the FK heterostructure are less sensitive to the bias voltage 
and do not show the resonance feature unlike the Hubbard heterostructure (see Fig.~\ref{fig:spectra_Hubb} for comparison). 
In the FK model, electric current is carried only by the itinerant $c$ fermions. 
Numerical results for the current-voltage characteristics of $N=6$ FK heterostructure are shown as filled circles in Fig.~\ref{fig:IV_FK}. 
For comparison, the current per spin as a function of bias voltage for $N=6$ Hubbard heterostructure with $U=15t$ is also shown as open circles. 
At first sight, the similarity in the $I$-$V$ characteristics is remarkable between the Hubbard and the FK heterostructures. 
This clearly demonstrates that the $z$-axis transport properties of metal--Mott-insulator--metal heterostructures is dominated by the inter-Hubbard-band tunneling. 
Yet, about 10\% (20\%) enhancement of the current at $eV \alt 15t$ $(eV \sim 17t)$ can be seen in the Hubbard heterostructure 
despite a factor 2/3 smaller transfer intensity than in the FK heterostructure. 
This originates from the faster narrowing of the Mott gap by an applied bias voltage in the Hubbard heterostructures than in the FK heterostructures. 
To see this behavior semiquantitatively, 
let us define the gap amplitude by the width in which the spectral function becomes smaller than 0.001. 
Note that the spectral function is finite everywhere and, strictly speaking, there is no gap at finite temperature and bias voltage. 
By this definition, the gap amplitude at $eV =14t$ is estimated to be $1.60t$ at layer $z=3$ in $N=6$ Hubbard heterostructure, 
and $2.66t$ at layer $z=3$ in $N=6$ FK heterostructure, 
i.e., the former is about 40\% smaller than the latter. 
Thus the effect of dynamical fluctuations in the Hubbard heterostructure is rather strong, 
overcoming the narrower bandwidth and enhancing the current amplitude. 

\begin{figure*}[tbp]
\includegraphics[width=1.4\columnwidth,clip]{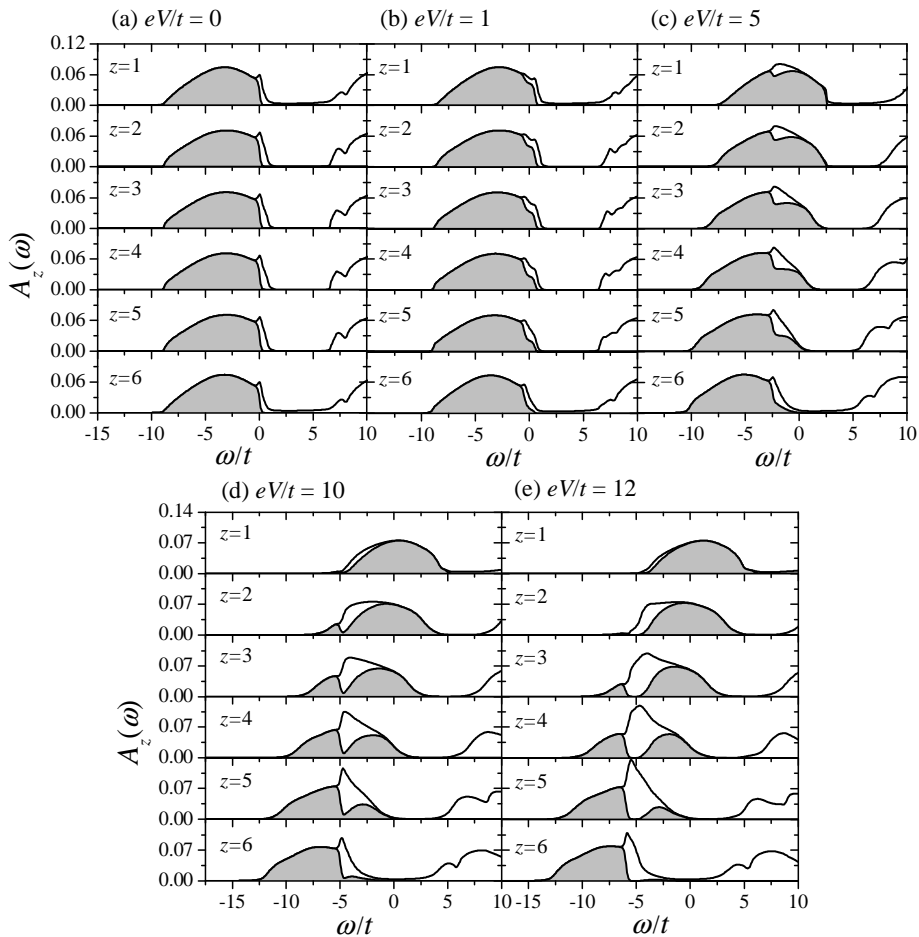}
\caption{Layer-resolved spectral function for $N=6$ doped Hubbard heterostructure with several choices of bias voltage indicated. 
On-site interaction is $U=15t$, on-site potentials at all layer are shifted by $4t$ while the chemical potential at equilibrium was fixed to 0. 
Thus some amount of holes is doped into the interacting region. 
Shaded area shows the region occupied by electrons. 
In (b), $eV=t$, two quasiparticlelike sharp structures are visible in $A_z(\omega)$ at $\omega=\pm eV/2$ corresponding to the chemical potentials of leads $L$ and $R$. 
On the other hand, in (c)--(e), only one structure can be seen at $\omega=-eV/2$.
The other one, which is expected at $\omega=eV/2$, is moved up inside the Mott gap and 
therefore becomes invisible except for a steplike feature seen at $z=1$ in (c).}
\label{fig:spectra_doped}
\end{figure*}

\begin{figure}[tbp]
\includegraphics[width=0.8\columnwidth,clip]{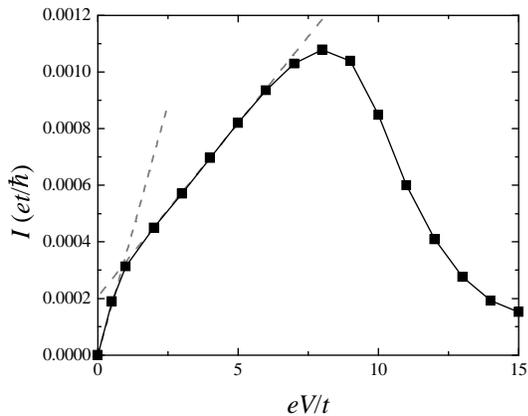}
\caption{Current-voltage characteristics of $N=6$ Hubbard heterostructure with $U=15t$ and on-site potential shifted by $4t$. }
\label{fig:IV_doped}
\end{figure}

\begin{figure}[tbp]
\includegraphics[width=0.8\columnwidth,clip]{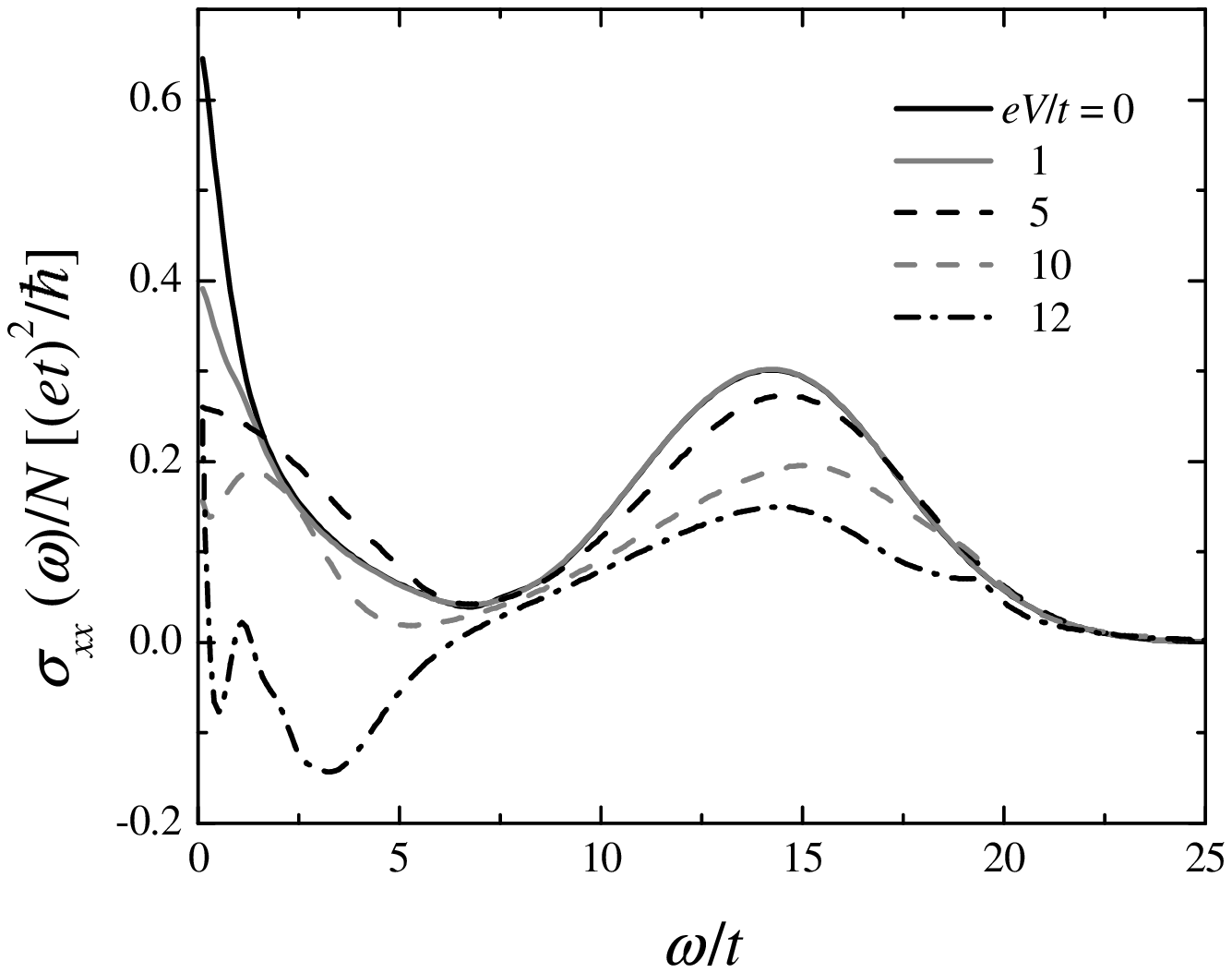}
\caption{Optical conductivity spectra of $N=6$ Hubbard heterostructure with several choices of bias voltage indicated. 
On-site interaction $U=15t$ and on-site potential shifted by $4t$. 
}
\label{fig:opcon_doped}
\end{figure}

\begin{figure}[tbp]
\includegraphics[width=0.8\columnwidth,clip]{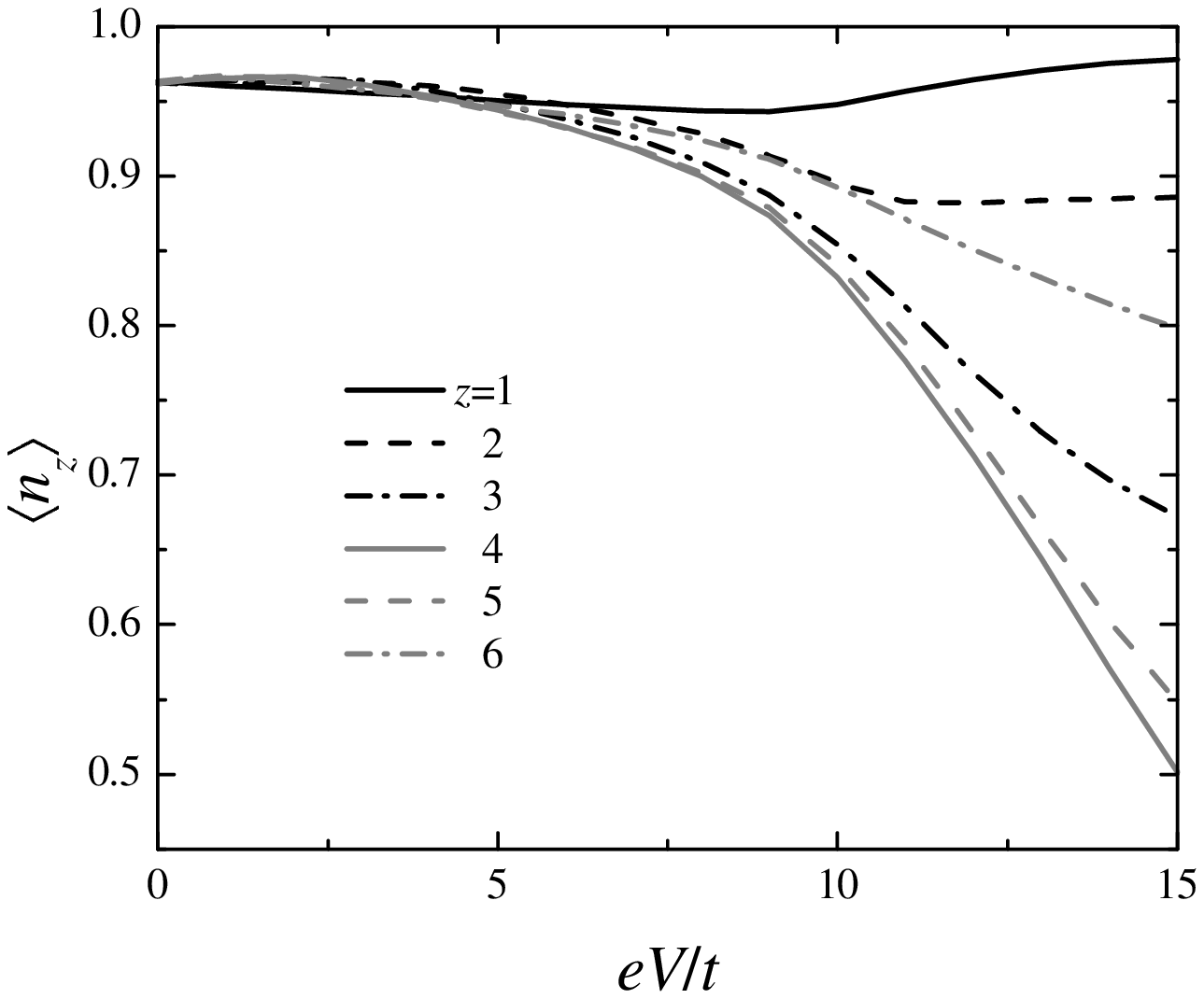}
\caption{Charge densities as functions of bias voltage for $N=6$ Hubbard heterostructure with $U=15t$ and on-site potential shifted by $4t$. 
 }
\label{fig:n_doped}
\end{figure}

\section{Summary and Discussion}
\label{sec:summary}
Summarizing, we have investigated the electronic properties of 
metal--Mott-insulator--metal heterostructures under an applied bias voltage between the metallic leads 
by employing the dynamical-mean-field theory combined with the Keldysh Green's function technique. 
We have focused on the strong coupling region. 
In this case, the current-voltage characteristics of Hubbard heterostrucures 
were found to be quite similar to that of the conventional metal-semiconductor-metal heterostructures. 
Similar current-voltage characteristics are also obtained by using the Falicov-Kimball model in which itinerant fermions do not exhibit a quasiparticle feature.\cite{Si92} 
These findings indicate that the electron transport in correlated-insulator heterostructures is mainly dominated by inter-Hubbard-band tunneling. 
On the other hand, the electron spectral functions are strongly deformed by an applied voltage. 
Deformed spectral functions under a finite bias voltage are measurable by using (spatially resolved) photoemission spectroscopy and scanning tunneling microscopy. 
Such effects may also be examined by optical conductivity measurements.

\subsection*{Doped case}

So far, we have considered the particle-hole symmetric situation before applying a bias voltage. 
Here we consider the hole-doped situation by increasing the potential at all layers by $4t$ with the other parameters unchanged. 
As in the undoped case, we assume that the potential in the central region varies linearly as 
$\varepsilon_0(z) = -U/2 + 4t + eV (N+1-2z)/(2N+2)$ interpolating $\varepsilon_L$ and $\varepsilon_R$. 
In a realistic situation, this assumption is not fully justified because an external electric field is screened by the redistribution of carriers and 
the potential drop occurs only near the interface regions. 
In order to describe such a screening effect, it is necessary to introduce a long-range Coulomb interaction 
and to determine the potential profile by solving the Poisson equation self-consistently. 
These effects are neglected in this study. 
Nevertheless, the qualitative behavior including the linear current-voltage characteristics and the change in quasiparticle features presented below 
is expected to be unchanged. 

Numerical results for the layer-resolved spectral functions are shown in Fig.~\ref{fig:spectra_doped}. 
At low bias voltage, the splitting of the resonance structure can be seen [Fig.~\ref{fig:spectra_doped} (b)]. 
This is similar to the nonequilibrium Kondo effect in the quantum dots.\cite{Meir91} 
With increasing $V$, higher peaks are moved inside the Mott gap, and only the lower peaks remain visible. 
Because the electric current is carried by the quaiparticles in the metallic region, 
the change in the spectral function by a bias voltage may affect the current-voltage characteristics. 

The current-voltage characteristic of the doped Hubbard heterostructure is shown in Fig.~\ref{fig:IV_doped}. 
Near-linear voltage dependence of current can be seen at low voltage $eV \alt 7t$ with 
the change of slope at $eV \sim t$ as indicated by dashed lines. 
This change is due to the disappearance of one of the two resonance peaks. 

The frequency dependent electron occupation in the doped Hubbard heterostructure is changed more drastically. 
As shown in Fig.~\ref{fig:spectra_doped} (e), holes are injected into the lower Hubbard band more strongly than in the undoped case 
(see Fig.~\ref{fig:spectra_Hubb} for comparison). 
Thus the electron density becomes completely depressed in some frequency regions (see layers at $z=3,4,5$). 
These features may affect excitations, i.e., the optical conductivity, more strongly than in the undoped case. 

Numerical results for the in-plane optical conductivity of the doped Hubbard heterostructure are shown in Fig.~\ref{fig:opcon_doped} 
for several choices of bias voltage indicated. 
Contrary to the Mott-insulator heterostructure, an applied bias voltage destroys the low frequency Drude-like peak efficiently. 
An applied bias voltage also reduces the total weight 
correlated with the reduction of electron density inside the interacting region as shown in Fig.~\ref{fig:n_doped}. 
The reduced electron density strongly depends on the layer, indicating the emergence of a ``dipole'' moment. 
Therefore, in a more realistic calculation, 
it will be necessary to include the long-ranged Coulomb interaction and determine the potential profile self-consistently.\cite{Okamoto04b} 
We are currently working on including such effects under various situations. 
The two-peak structure in the spectral function seen in Fig.~\ref{fig:spectra_doped} (b) is hardly observable in the optical conductivity. 
This may be because the electron distribution functions $f_{eff,z}$ have a similar structure to the spectral functions, 
effectively reducing the contribution from the interpeak transition.  
As can be seen from the shaded area, $f_{eff,z}$ has a two-step structure. 
Further, the finite imaginary part of the self-energy broadens the structure in the optical conductivity even if it exists. 

Another interesting observation is the emergence of the negative conductivity $\sigma_{xx} (\omega)<0$ 
(see the result for $eV=12t$ in Fig.~\ref{fig:opcon_doped}). 
The negative conductivity should be remedied by including the vertex correction. 
Yet, it implies the optical response of the correlated system becomes unusual in the nonequilibrium situation.  
This may be an interesting future problem.

\subsection*{Other work and future prospects}

Nonequilibrium transport properties of one-dimensional metal--Mott-insulator-metal heterostructures have also been studied by Yonemitsu.\cite{Yonemitsu05} 
His main focus is on the field-effect carrier injection by an applied gate voltage, and 
he points out that the ambipolar current-voltage characteristics is a general feature of metal--Mott-insulator-metal heterostructures. 
In contrast to the behavior of conventional metal--band-insulator--metal heterostructures, 
ambipolar behavior is found to be insensitive to the difference in the work function between the Mott insulator and the metallic leads. 
On the other hand, in light of our results, Mott insulators and band insulators are expected to behave quite similarly. 
Thus the difference comes from the absence of long-range Coulomb interaction and the difference in the work function between the sample and leads in our case. 
Including these effects is certainly necessary for applying the current DMFT method to the more realistic system. 
Yet, there may appear further interesting phenomena in higher dimension; 
in this case, quasiparticle structure at the Fermi level emerges more easily than in the particle-hole symmetric case studied here. 

Another work on one-dimensional heterostructures was performed by Oka and Nagaosa.\cite{Oka05} 
They applied the DMRG technique to study the charge density redistribution in 
the interface between noninteracting metal and correlated-metal in the presence of work function difference between the two. 
Even though they are dealing with one-dimensional models in which quantum effects are strongest, 
density profiles were found to be well reproduced by using the classical charge with the appropriate gap at an integer-filling region. 
This observation about the static properties 
is actually consistent with the previous studies on the higher-dimensional heterostructures. 
It has been shown that the static charge profile computed by DMFT is 
almost identical to the one by the mean-field approximation.\cite{Freericks01,Okamoto04b} 
Based on this finding, 
they proposed an interesting ``interface Mott transition'' to explain the colossal electroresistance, 
i.e., the large switching of resistance by an applied bias voltage, observed experimentally.\cite{Bikalov03,Sawa04} 
Their assumption is that the filling-controlled Mott metal-insulator transition occurs by applying a bias voltage, 
and that the Mott transition is of the first order in higher-than-one dimensions. 
The Mott metal-insulator transition in the higher dimension is characterized by a collapse of the small energy scale which is roughly 
the quasiparticle bandwidth.\cite{Georges96} 
Therefore, in order to have a first order transition, the temperature must be smaller than this energy scale. 
On the other hand, applying a bias voltage $V$ has a similar effect to increasing the temperature.  
Therefore if the bias voltage required to accomplish the integer filling at some layer 
becomes larger than the small energy scale, 
the metal-insulator transition would become of the second order or just a crossover. 
A very interesting theoretical question is whether an interface Mott transition is really possible. 

In this paper, we have focused only on a paramagnetic phase and ``Mott'' physics. 
However in general, correlated electron systems have a strong tendency toward magnetic ordering. 
A single-band Hubbard model is known to exhibit an antiferromagnetic ordering at half-filling. 
This ordering is affected by applying a magnetic field, changing the chemical potentials by a gate voltage, or injecting a current by applying a bias voltage. 
In any of these cases, the current-voltage characteristics are expected to be modified from the results obtained in this paper. 
Therefore including magnetic symmetry breaking is one of the interesting extensions of this work. 
It is also possible and desirable to consider magnetic leads instead of non-magnetic ones. 
This is of the direct relevance to the TMR effect. 

Further, application of the present DMFT method to other models is highly desirable. 
This includes the double-exchange model (with electron-phonon coupling) for CMR manganites, 
and the multiorbital Hubbard model for general transition-metal oxides. 
For either model, fluctuation, ordering or melting of internal degrees of freedom would affect the transport properties of heterostructures. 
Controlling these degrees of freedom by external fields including a bias voltage and exploring the new phenomena that arise are 
important and urgent tasks. 

Another possible extension is including spatial correlation beyond single-site DMFT. 
As the mathematical structure of the EOM method used in this paper is quite similar to that of so called 
correlator projection method,\cite{Onoda03} 
generalizing the current method to the multisite problem would not be so difficult. 

\acknowledgements
The author thanks 
E. Dagotto, K. A. Al-Hassanieh, R. S. Fishman, X.-G Zhang, A. H. Castro Neto,
J. K. Freericks, J. E. Han, A. J. Millis, and I. Zutic 
for valuable discussions 
and R. S. Fishman for critical reading of the manuscript. 
This work was supported by the Division of Materials Sciences and Engineering, Office of Basic Energy Sciences, U.S. Department of Energy, 
under Contract No. DE-AC05-00OR22725 with Oak Ridge National Laboratory, managed and operated by UT-Battelle, LLC.

\end{document}